\documentclass[runningheads]{llncs}
\usepackage[colorlinks]{hyperref}
\usepackage{amsmath,amssymb,mathtools}
\usepackage{stmaryrd}
\usepackage{wasysym}
\usepackage{graphicx}
\usepackage{enumerate}
\usepackage{tikz}
\usetikzlibrary{arrows}
\usetikzlibrary{arrows.meta}
\usetikzlibrary{calc,decorations.pathmorphing,shapes}
\usepackage[all]{xy} 
\usepackage{dsfont}
\numberwithin{equation}{section}

\newtheorem{notation}{Notation}

\newcommand{\be}{\begin{equation}}
\newcommand{\ee}{\end{equation}}
\newcommand{\bprf}{\begin{proof}}
\newcommand{\eprf}{\end{proof}}

\newcommand\define[1]{\emph{\textbf{#1}}}

\newcommand{\id}{\mathrm{id}}
\newcommand{\tr}{{\rm tr} }
\newcommand{\Ad}{\mathrm{Ad}}
\newcommand{\RE}{\mathrm{RE}}
\let\ov=\overline

\def\mA{\mathcal{A}}
\def\mB{\mathcal{B}}
\def\mC{\mathcal{C}}

\def\mM{\mathcal{M}}

\def\B{\mathbb{B}}
\def\C{\mathbb{C}}

\def\N{\mathbb{N}}

\DeclarePairedDelimiterX{\relentx}[2]{(}{)}{%
  #1\;\delimsize\|\;#2%
}
\newcommand{\relent}{S\relentx}

\newcommand{\stoch}{\;\xy0;/r.25pc/:(-3,0)*{}="1";(3,0)*{}="2";{\ar@{~>}"1";"2"|(1.06){\hole}};\endxy\!}
\newcounter{sarrow}
\newcommand\xstoch[1]{%
\stepcounter{sarrow}%
\mathrel{\begin{tikzpicture}[baseline= {( $ (current bounding box.south) + (0,-0.1ex) $ )}]
\node[inner sep=.5ex] (\thesarrow) {\;$\scriptstyle #1$\;};
\path[draw,{<[scale=1.5,width=3,length=2]}-,decorate,
  decoration={snake,amplitude=0.3mm,segment length=2.1mm,pre=lineto,pre length=1pt}] 
    (\thesarrow.south east) -- (\thesarrow.south west);
\end{tikzpicture}}%
}
\newcommand{\too}{\xy0;/r.25pc/:(-3,0)*{}="1";(3,0)*{}="2";{\ar@{~>}@<-0.5ex>"2";"1"|(1.06){\hole}};{\ar@<-0.5ex>"1";"2"};\endxy\;}
\newcommand\rt{\rightarrowtriangle}

\providecommand\longrightarrowRHD{\relbar\joinrel\relbar\joinrel\mathrel\RHD}
\makeatletter
\providecommand*\xrt[2][]{%
  \ext@arrow 0055{\arrowfill@\relbar\relbar\rightarrowtriangle}{#1}{#2}}
  \providecommand*\xrtb[2][]{\ext@arrow 0055{\arrowfill@\relbar\relbar\longrightarrowRHD}{#1}{#2}}
\makeatother

\newcommand{\bigboxplus}{
  \mathop{
    \vphantom{\bigoplus} 
    \mathchoice
      {\vcenter{\hbox{\resizebox{\widthof{$\displaystyle\bigoplus$}}{!}{$\boxplus$}}}}
      {\vcenter{\hbox{\resizebox{\widthof{$\bigoplus$}}{!}{$\boxplus$}}}}
      {\vcenter{\hbox{\resizebox{\widthof{$\scriptstyle\oplus$}}{!}{$\boxplus$}}}}
      {\vcenter{\hbox{\resizebox{\widthof{$\scriptscriptstyle\oplus$}}{!}{$\boxplus$}}}}
  }\displaylimits 
}

\newcommand{\NCFinStat}{\mathbf{NCFinStat}}
\newcommand{\FinStat}{\mathbf{FinStat}}
\newcommand{\NCFP}{\mathbf{NCFP}}
\newcommand{\FP}{\mathbf{FP}}
\newcommand{\op}{\mathrm{op}}

\begin{document}
\emergencystretch 2em
%
\title{Towards a functorial description of quantum relative entropy}
%
%
\author{Arthur J.\ Parzygnat\orcidID{0000-0002-7264-3991}}
\authorrunning{A.\ Parzygnat}
%
\institute{Institut des Hautes \'Etudes Scientifiques, 35 Route de Chartres 91440, Bures-sur-Yvette, France
\email{parzygnat@ihes.fr}
}
\maketitle              
\begin{abstract}
A Bayesian functorial characterization of the classical relative entropy (KL divergence) of finite probabilities was recently obtained by Baez and Fritz. This was then generalized to standard Borel spaces by Gagn\'e and Panangaden. 
Here, we provide preliminary calculations suggesting that the finite-dimensional quantum (Umegaki) relative entropy might be characterized 
in a similar way. Namely, we explicitly prove that it defines an affine functor in the special case where the relative entropy is finite. A recent non-commutative disintegration theorem provides a key ingredient in this proof. 

\keywords{Bayesian inversion \and disintegration \and optimal hypothesis.}
\end{abstract}
%
%
%
\section{Introduction and outline}

In 2014, Baez and Fritz provided a categorical Bayesian characterization of the relative entropy of finite probability measures using a category of hypotheses~\cite{BaFr14}. This was then extended to standard Borel spaces by Gagn\'e and Panangaden in 2018~\cite{GaPa18}. An immediate question remains as to whether or not the quantum (Umegaki) relative entropy~\cite{Um62} has a similar characterization.%
\footnote{The ordinary Shannon and von~Neumann entropies were characterized in~\cite{BFL} and~\cite{PaEntropy}, respectively, in a similar categorical setting.}
The purpose of the present work is to begin filling this gap by using the recently proved non-commutative disintegration theorem~\cite{PaRu19}. 

The original motivation of Baez and Fritz came from Petz' characterization of the quantum relative entropy~\cite{Pe92}, which used a quantum analogue of hypotheses known as conditional expectations. 
Although Petz' characterization had some minor flaws, which were noticed in~\cite{BaFr14}, we believe Petz' overall idea is correct when formulated on an appropriate category of non-commutative probability spaces and non-commutative hypotheses. 
In this article, we show how the Umegaki relative entropy defines an affine functor that vanishes on the subcategory of non-commutative optimal hypotheses for faithful states. 
The chain rule for quantum conditional entropy is a consequence of functoriality. 
The non-faithful case 
will be addressed in future work, where we hope to provide a characterization of the quantum relative entropy as an affine functor.

\section{The categories of hypotheses and optimal hypotheses}

In this section, we introduce non-commutative analogues of the categories from~\cite{BaFr14}. All $C^*$-algebras here are finite-dimensional and unital. All $*$-homomorphisms are unital unless stated otherwise. $\mM_{n}$ denotes the algebra of $n\times n$ matrices. If $V:\C^{m}\to\C^{n}$ is a linear map, $\Ad_{V}:\mM_{n}\stoch\mM_{m}$ denotes the linear map sending $A$ to $VAV^{\dag}$, where $V^{\dag}$ is the adjoint (conjugate transpose) of $V$. Linear maps between algebras are written with squiggly arrows $\stoch$, while $*$-homomorphisms are written as straight arrows $\rightarrow$. The acronym CPU stands for ``completely positive unital.'' 
If $\mA$ and $\mB$ are matrix algebras, then $\tr_{\mA}:\mA\otimes\mB\stoch\mB$ denotes the \define{partial trace over $\mA$} and is the unique linear map determined by $\tr_{\mA}(A\otimes B)=\tr(A)B$ for $A\in\mA$ and $B\in\mB$. If $\omega$ is a state on $\mA$, its quantum entropy is 
denoted by $S(\omega)$ 
(cf.\ \cite[Definition~2.20]{PaEntropy}).

\begin{definition}
Let $\NCFinStat$ be the category of \define{non-commutative probability spaces}, whose objects are pairs $(\mA,\omega)$, with $\mA$ a $C^*$-algebra and $\omega$ a state on $\mA$. A morphism $(\mB,\xi)\rt(\mA,\omega)$ is a pair $(F,Q)$ with $F:\mB\to\mA$ a $*$-homomorphism and $Q:\mA\stoch\mB$ a CPU map (called  a \define{hypothesis}), such that
\[
\omega\circ F=\xi
\qquad\text{ and }\qquad
Q\circ F=\id_{\mB}.
\]
The composition rule in $\NCFinStat$ is given by 
\[
(\mC,\zeta)\xrt{(G,R)}(\mB,\xi)\xrt{(F,Q)}(\mA,\omega)
\quad\mapsto\quad
(\mC,\zeta)\xrt{(F\circ G,R\circ Q)}(\mA,\omega).
\]
Let $\NCFP$ be the subcategory of $\NCFinStat$ with the same objects but whose morphisms are pairs $(F,Q)$ as above and $Q$ is an \define{optimal hypothesis}, 
i.e.\ $\xi\circ Q=\omega$.
\end{definition}

The subcategories of $\NCFinStat^{\op}$ and $\NCFP^{\op}$ consisting of \emph{commutative} $C^*$-algebras are equivalent to the categories $\FinStat$ and $\FP$ from~\cite{BaFr14} by stochastic Gelfand duality (cf.\ \cite[Sections~2.5 and~2.6]{Pa17},~\cite{FuJa15}, and~\cite[Corollary~3.23]{PaRu19}).

\begin{notation}
\label{not:ABC}
On occasion, 
the notation $\mA$, $\mB$, and $\mC$ will be used to mean 
\[
\mA:=\bigoplus_{x\in X}\mM_{m_{x}},\qquad
\mB:=\bigoplus_{y\in Y}\mM_{n_{y}},\quad
\text{ and }\quad
\mC:=\bigoplus_{z\in Z}\mM_{o_{z}},
\]
where $X,Y,Z$ are finite sets, often taken to be ordered sets $X=\{1,\dots,s\}$, $Y=\{1,\dots,t\}$, $Z=\{1,\dots,u\}$ for convenience (cf.\ \cite[Section~5]{PaRu19} and/or \cite[Example~2.2]{PaEntropy}). 
Note that every element of $\mA$ (and analogously for $\mB$ and $\mC$) is of the form $A=\bigoplus_{x\in X}A_{x}$, with $A_{x}\in\mM_{m_{x}}$. 
Furthermore, $\omega$, $\xi$, and $\zeta$ will refer to states on $\mA$, $\mB$, and $\mC$, respectively, with decompositions of the form
\[
\omega=\sum_{x\in X}p_{x}\tr(\rho_{x}\;\cdot\;),\qquad
\xi=\sum_{y\in Y}q_{y}\tr(\sigma_{y}\;\cdot\;),\quad
\text{ and }\quad
\zeta=\sum_{z\in Z}r_{z}\tr(\tau_{z}\;\cdot\;). 
\]
If $Q:\mA\stoch\mB$ is a linear map, its $yx$ component $Q_{yx}$ is the linear map obtained from the composite $\mM_{m_{x}}\hookrightarrow\mA\xstoch{Q}\mB\twoheadrightarrow\mM_{n_{y}}$, where the first and last maps are the (non-unital) inclusion and projection, respectively. 
\end{notation}

\begin{definition}
\label{defn:standardform}
Let $\mA$ and $\mB$ be as in Notation~\ref{not:ABC}. A morphism $(\mB,\xi)\xrt{(F,Q)}(\mA,\omega)$ in $\NCFinStat$ is said to be in \define{standard form} iff there exist non-negative integers $c^{F}_{yx}$ such that (cf.\ \cite[Lemma~2.11]{PaEntropy} and \cite[Theorem~5.6]{Fa01})
\[
F(B)=\bigoplus_{x\in X}\bigboxplus_{y\in Y}\big(\mathds{1}_{c^{F}_{yx}}\otimes B_{y}\big)\equiv\bigoplus_{x\in X}\mathrm{diag}\left(\mathds{1}_{c^{F}_{1x}}\otimes B_{1},\dots,\mathds{1}_{c^{F}_{tx}}\otimes B_{t}\right)
\quad\forall\;B\in\mB, 
\]
which is a direct sum of block diagonal matrices. 
The number $c^{F}_{yx}$ is called the \define{multiplicity} of $\mM_{n_{y}}$ in $\mM_{m_{x}}$ associated to $F$. 
In this case, each $A_{x}\in\mM_{m_{x}}$ will occasionally be decomposed as $A_{x}=\sum_{y,y'\in Y}E_{yy'}^{(t)}\otimes A_{x;yy'}$, where $\{E_{yy'}^{(t)}\}$ denote the matrix units of $\mM_{t}$ and $A_{x;yy'}$ is a $(c^{F}_{yx}n_{y})\times(c^{F}_{y'x}n_{y'})$ matrix. 
\end{definition}

If $F$ is in standard form and if $\omega$ and $\xi$ are states on $\mA$ and $\mB$ (as in Notation~\ref{not:ABC}) such that $\xi=\omega\circ F$, then (cf.\ \cite[Lemma~2.11]{PaEntropy} and \cite[Proposition~5.67]{PaRu19})
\be
\label{eq:sigmayintermsofrhox}
q_{y}\sigma_{y}=\sum_{x\in X}p_{x}\tr_{\mM_{c^{F}_{yx}}}(\rho_{x;yy}).
\ee 
The standard form of a morphism will be useful later for proving functoriality of relative entropy, and it will allow us to formulate expressions more explicitly in terms of matrices. 

\begin{lemma}
\label{lem:rectifytostandardformA}
Given a morphism $(\mB,\xi)\xrt{(F,Q)}(\mA,\omega)$
in $\NCFinStat$, with $\mA$ and $\mB$ be as in Notation~\ref{not:ABC}, there exists a unitary 
$U\in\mA$ such that
$(\mB,\xi)\xrt{(\Ad_{U^{\dag}}\circ F,Q\circ\Ad_{U})}(\mA,\omega\circ\Ad_{U})$
is a morphism in $\NCFinStat$ that is in standard form. Furthermore, if $(F,Q)$ is in $\NCFP$, then $(\Ad_{U^{\dag}}\circ F,Q\circ\Ad_{U})$ is also in $\NCFP$. 
\end{lemma}

\bprf
First, $(\Ad_{U^{\dag}}\circ F,Q\circ\Ad_{U})$ is in $\NCFinStat$ for \emph{any} unitary $U$ because 
\[
(\omega\circ\Ad_{U})\circ(\Ad_{U^{\dag}}\circ F)=\xi
\quad\text{ and }\quad
(Q\circ\Ad_{U})\circ(\Ad_{U^{\dag}}\circ F)=\id_{\mB},
\]
so that the two required conditions hold. Second, the fact that a unitary $U$ exists such that $F$ is in the form in Definition~\ref{defn:standardform} is a standard fact regarding (unital) $*$-homomorphisms between direct sums of matrix algebras~\cite[Theorem~5.6]{Fa01}. Finally, if $(F,Q)$ is in $\NCFP$, which means $\xi\circ Q=\omega$, then $(\Ad_{U^{\dag}}\circ F,Q\circ\Ad_{U})$ is also in $\NCFP$ because $\xi\circ(Q\circ\Ad_{U})=(\xi\circ Q)\circ\Ad_{U}=\omega\circ\Ad_{U}$. 
\eprf

Although the composite of two morphisms in standard form is \emph{not} necessarily in standard form, a permutation can always be applied to obtain one. Furthermore, a pair of composable morphisms in $\NCFinStat$ can also be simultaneously rectified. 
This is the content of the following lemmas.  

\begin{lemma}
\label{lem:rectifytostandardformB}
Given a composable pair $(\mC,\zeta)\!\xrt{\!\!(G,R)\!\!}\!(\mB,\xi)\!\xrt{\!\!(F,Q)\!\!}\!(\mA,\omega)$
in $\NCFinStat$, there exist unitaries 
$U\in\mA$ and $V\in\mB$ such that 
\[
(\mC,\zeta)\xrt{(\Ad_{V^{\dag}}\circ G,R\circ\Ad_{V})}(\mB,\xi\circ\Ad_{V})\xrt{(\Ad_{U^{\dag}}\circ F\circ\Ad_{V},\Ad_{V^{\dag}}\circ Q\circ\Ad_{U})}(\mA,\omega\circ\Ad_{U})
\]
is a pair of composable morphisms in $\NCFinStat$ that are \emph{both} in standard form.  
\end{lemma}

\bprf
By Lemma~\ref{lem:rectifytostandardformA}, there exists a unitary $V\in\mB$ such that 
\[
(\mC,\zeta)\xrt{(\Ad_{V^{\dag}}\circ G,R\circ\Ad_{V})}(\mB,\xi\circ\Ad_{V})\xrt{(F\circ\Ad_{V},\Ad_{V^{\dag}}\circ Q)}(\mA,\omega)
\]
is a composable pair of morphisms in $\NCFinStat$ with the left morphism in standard form. The right 
morphism is indeed 
in $\NCFinStat$ because 
\[
\omega\circ(F\circ\Ad_{V})=(\omega\circ F)\circ\Ad_{V}=\xi\circ\Ad_{V}
\qquad\text{ and }
\]
\[
(\Ad_{V^{\dag}}\circ Q)\circ(F\circ\Ad_{V})=\Ad_{V^{\dag}}\circ (Q\circ F)\circ\Ad_{V}=\Ad_{V^{\dag}}\circ\Ad_{V}=\id_{\mB}.
\] 
Then, applying Lemma~\ref{lem:rectifytostandardformA} again, but to the \emph{new} morphism on the right, gives a unitary $U$ satisfying the conditions claimed. 
\eprf

\begin{lemma}
\label{lem:pairinstandardform}
Given a composable pair $(\mC,\zeta)\!\xrt{\!\!(G,R)\!\!}\!(\mB,\xi)\!\xrt{\!\!(F,Q)\!\!}\!(\mA,\omega)$
in $\NCFinStat$, each in standard form, there exist \emph{permutation} matrices $P_{x}\in\mM_{m_{x}}$ such that 
\[
(\mC,\zeta)\xrt{(\Ad_{P^{\dag}}\circ F\circ G,R\circ Q\circ\Ad_{P})}(\mA,\omega\circ\Ad_{P})
\]
is also in standard form, where $P:=\bigoplus_{x\in X}P_{x}$ and the multiplicities $c^{G\circ F}_{zx}$ of $\mathrm{Ad}_{P^{\dag}}\circ G\circ F$ are given by 
$c^{G\circ F}_{zx}=\sum_{y\in Y}c^{G}_{zy}c^{F}_{yx}.$
\end{lemma}

\bprf
The composite $F\circ G$ is given by 
\[
F\big(G(C)\big)
=F\Bigg(\bigoplus_{y\in Y}\underbrace{\bigboxplus_{z\in Z}\big(\mathds{1}_{c^{G}_{yz}}\otimes C_{z}\big)}_{B_{y}}\Bigg)
=\bigoplus_{x\in X}\underbrace{\bigboxplus_{y\in Y}\Bigg(\mathds{1}_{c^{F}_{yx}}\otimes\bigboxplus_{z\in Z}\big(\mathds{1}_{c^{G}_{yz}}\otimes C_{z}\big)\Bigg)}_{A_{x}}.
\]
The matrix $A_{x}$ takes the more explicit form (with zeros in unfilled entries)
\[
A_{x}=\mathrm{diag}\left(\!\mathds{1}_{c^{F}_{1x}}\otimes\left[\begin{smallmatrix}\mathds{1}_{c^{G}_{11}}\otimes C_{1}&&\\&\ddots&\\&&\mathds{1}_{c^{G}_{u1}}\otimes C_{u}\end{smallmatrix}\right],\dots,\mathds{1}_{c^{F}_{tx}}\otimes\left[\begin{smallmatrix}\mathds{1}_{c^{G}_{11}}\otimes C_{1}&&\\&\ddots&\\&&\mathds{1}_{c^{G}_{u1}}\otimes C_{u}\end{smallmatrix}\right]\!\right).
\] 
From this, one sees that the number of times $C_{z}$ appears on the diagonal is $\sum_{y\in Y}c^{G}_{zy}c^{F}_{yx}$. 
However, the positions of $C_{z}$ are not all next to each other. Hence, a permutation matrix $P_{x}$ is needed to put them into standard form. 
\eprf

\begin{notation}
\label{not:pullbackstates}
Given a composable pair $(\mC,\zeta)\xrt{(G,R)}(\mB,\xi)\xrt{(F,Q)}(\mA,\omega)$ in standard form as in Lemma~\ref{lem:pairinstandardform}, the states $\zeta\circ R$ and $\xi\circ Q$ will be decomposed as
\[
\zeta\circ R=\sum_{y\in Y}q_{y}^{R}\tr\big(\sigma_{y}^{R}\;\cdot\;\big)
\qquad\text{ and }\qquad
\xi\circ Q=\sum_{x\in X}p_{x}^{Q}\tr\big(\rho_{x}^{Q}\;\cdot\;\big).
\]
\end{notation}

\begin{lemma}
\label{lem:explicitformstandard}
Given a morphism $(F,Q)$ in standard form as in Notation~\ref{not:pullbackstates} such that all states are faithful, 
 there exist 
strictly positive matrices 
$\alpha_{yx}\in\mM_{c^{F}_{yx}}$
for all $x\in X$ and $y\in Y$ such that 
\[
\tr\left(\sum_{x\in X}\alpha_{yx}\right)=1\quad\forall\;y\in Y, \qquad
p_{x}^{Q}\rho_{x}^{Q}=\bigboxplus_{y\in Y}(\alpha_{yx}\otimes q_{y}\sigma_{y})
\;\;\forall\;x\in X, \quad\text{and}
\]
\[
Q_{yx}(A_{x})=\tr_{\mM_{c_{yx}^{F}}}\big((\alpha_{yx}\otimes\mathds{1}_{n_{y}})A_{x;yy}\big)
\qquad\forall\;y\in Y
,\; A_{x}\in\mM_{m_{x}},\;x\in X.
\]
\end{lemma}

\bprf
Because $Q$ and $R$ are disintegrations of $(F,\xi\circ Q,\xi)$ and $(G,\zeta\circ R,\zeta)$, respectively, the claim follows from the non-commutative disintegration theorem~\cite[Theorem~5.67]{PaRu19} and the fact that $F$ is an injective $*$-homomorphism. The $\alpha_{yx}$ matrices are strictly positive by the faithful assumption. 
\eprf

If $(\mC,\zeta)\xrt{(G,R)}(\mB,\xi)\xrt{(F,Q)}(\mA,\omega)$ is composable pair, a consequence of 
Lemma~\ref{lem:explicitformstandard} is 
\be
\label{eq:zRQ}
\zeta\circ R\circ Q=\sum_{x\in X}\tr\Bigg(\Bigg(\bigboxplus_{y\in Y}\alpha_{yx}\otimes q_{y}^{R}\sigma^{R}_{y}\Bigg)\;\cdot\;\Bigg).
\ee

\section{The relative entropy as a functor}

\begin{definition}
\label{defn:relent}
Set $\RE:\NCFinStat\rightarrow\B(-\infty,\infty]$ to be the assignment that sends a morphism $(\mB,\xi)\xrt{(F,Q)}(\mA,\omega)$ to $\relent{\omega}{\xi\circ Q}$
(the assignment is trivial on objects). Here, $\B M$ is the one object category associated to any monoid%
\footnote{The morphisms of $\B M$ from that single object to itself equals the set $M$ and the composition is the monoid multiplication. Here, the monoid is $(-\infty,\infty]$ under addition (with the convention that $a+\infty=\infty$ for all $a\in(-\infty,\infty]$.}
 $M$, $\relent{\cdot}{\cdot}$ is the \define{relative entropy} of two states on the same $C^*$-algebra, which is defined on an ordered pair of states $(\omega,\omega')$, with $\omega\preceq\omega'$ (meaning $\omega'(a^*a)=0$ implies $\omega(a^*a)=0$), on $\mA=\bigoplus_{x\in X}\mM_{m_{x}}$ by
\[
\relent{\omega}{\omega'}:=\tr\left(\bigoplus_{x\in X}p_{x}\rho_{x}\Big(\log(p_{x}\rho_{x})-\log(p'_{x}\rho'_{x})\Big)\right), 
\]
where $0\log0:=0$ by convention. If $\omega\npreceq\omega'$, then $\relent{\omega}{\omega'}:=\infty$.  
\end{definition}

\begin{lemma}
\label{lem:REproperties}
Using the notation from Definition~\ref{defn:relent}, the following facts hold. 
\begin{enumerate}[(a)]
\item
$\RE$ factors through $\B[0,\infty]$.
\item
$\RE$ vanishes on the subcategory $\NCFP$. 
\item
$\RE$ is invariant with respect to changing a morphism to standard form, i.e.\ 
in terms of the notation introduced in Lemma~\ref{lem:rectifytostandardformA}, 
\[
\RE\left((\mB,\xi)\xrt{(F,Q)}(\mA,\omega)\right)
=
\RE\left((\mB,\xi)\xrt{(\Ad_{U^{\dag}}\circ F,Q\circ\Ad_{U})}(\mA,\omega\circ\Ad_{U})\right)
.
\]
\end{enumerate}
\end{lemma}

\bprf
Left as an exercise.
\eprf

\begin{proposition}
\label{prop:somefunctoriality}
For a composable pair $(\mC,\zeta)\xrt{(G,R)}(\mB,\xi)\xrt{(F,Q)}(\mA,\omega)$
in $\NCFinStat$ (with all states and CPU maps faithful),%
\footnote{Faithfulness guarantees the finiteness of all expressions. More generally, our proof works if the appropriate absolute continuity conditions hold. Also, note that the ``conditional expectation property'' in~\cite{Pe92} is a special case of functoriality applied to a composable pair of morphisms of the form $(\C,\id_{\C})\xrt{(!_{\mB},R)}(\mB,\xi)\xrt{(F,Q)}(\mA,\omega)$, where $!_{\mB}:\C\to\mB$ is the unique unital linear map. Indeed, Petz' $\mA$, $\mB$, $E$, $\omega_{|\mA}$, $\varphi_{|\mA}$, and $\varphi$, are our $\mB$, $\mA$, $Q$, $\xi$, $R$, and $R\circ Q$, respectively ($\omega$ is the same).}
$
\relent{\omega}{\zeta\circ R\circ Q}=\relent{\xi}{\zeta\circ R}+\relent{\omega}{\xi\circ Q}, 
$
i.e.\
$\RE\big((F\circ G,R\circ Q)\big)=\RE\big((G,R)\big)+\RE\big((F,Q)\big)$. 
\end{proposition}

\bprf
By Lemma~\ref{lem:REproperties}, it suffices to assume $(F,Q)$ and $(G,R)$ are in standard form. To prove the claim, we 
expand each term. 
First,%
\footnote{Equation~(\ref{eq:SomxQ}) is a generalization of Equation~(3.2) in~\cite{BaFr14}, which plays a crucial role in proving many claims. We will also use it to prove affinity of $\RE$.}
\be
\label{eq:SomxQ}
\begin{split}
\relent{\omega}{\xi\circ Q}
&\overset{\text{Lem~\ref{lem:explicitformstandard}}}{=\joinrel=\joinrel=\joinrel=}
-S(\omega)-
\sum_{x\in X}\tr\Bigg(p_{x}\rho_{x}\log\Bigg(\bigboxplus_{y\in Y}\alpha_{yx}\otimes q_{y}\sigma_{y}\Bigg)\Bigg)\\
&=-S(\omega)-\sum_{x\in X}\sum_{y\in Y}\tr\Big(p_{x}\rho_{x;yy}\big(\log(\alpha_{yx})\otimes\mathds{1}_{n_{y}}\big)\Big)\\
&\quad-\sum_{x\in X}\sum_{y\in Y}\tr\Big(p_{x}\tr_{\mM_{c^{F}_{yx}}}(\rho_{x;yy})\log(q_{y}\sigma_{y})\Big).
\end{split}
\ee
The last equality follows from the properties of the trace, partial trace, and logarithms of tensor products.
By similar arguments, 
\be
\label{eq:SxzR}
\begin{split}
\relent{\xi}{\zeta\circ R}
&\overset{\text{(\ref{eq:sigmayintermsofrhox})}}{=\joinrel=\joinrel=}\sum_{x\in X}\sum_{y\in Y}\tr\Big(p_{x}\tr_{\mM_{c^{F}_{yx}}}(\rho_{x;yy})\log(q_{y}\sigma_{y})\Big)\\
&\quad-\sum_{x\in X}\sum_{y\in Y}\tr\Big(p_{x}\tr_{\mM_{c^{F}_{yx}}}(\rho_{x;yy})\log(q^{R}_{y}\sigma^{R}_{y})\Big)
\end{split}
\ee
and
\be
\label{eq:SomzRQ}
\begin{split}
\relent{\omega}{\zeta\circ R\circ Q}&\overset{\text{(\ref{eq:zRQ})}}{=\joinrel=\joinrel=}
-S(\omega)-\sum_{x\in X}\sum_{y\in Y}\tr\left(p_{x}\rho_{x;yy}\Big(\log(\alpha_{yx})\otimes\mathds{1}_{n_{y}}\Big)\right)\\
&\quad-\sum_{x\in X}\sum_{y\in Y}\tr\Big(p_{x}\tr_{\mM_{c^{F}_{yx}}}(\rho_{x;yy})\log(q^{R}_{y}\sigma^{R}_{y})\Big).
\end{split}
\ee
Hence, $(\ref{eq:SomxQ})+(\ref{eq:SxzR})=(\ref{eq:SomzRQ})$, which proves the claim. 
\eprf

\begin{example}
The usual chain rule for the quantum conditional entropy is a special case of Proposition~\ref{prop:somefunctoriality}. 
To see this, set $\mA:=\mM_{d_{A}},\mB:=\mM_{d_{B}},\mC:=\mM_{d_{C}}$ with $d_{A},d_{B},d_{C}\in\N$. 
Given a density matrix $\rho_{ABC}$ on $\mA\otimes\mB\otimes\mC$, we implement subscripts to denote the associated density matrix after tracing out a subsystem.
The chain rule for the conditional entropy states
\be
\label{eq:qchainrule}
H(AB|C)=H(A|BC)+H(B|C),
\ee
where (for example)
\[
H(B|C):=\tr(\rho_{BC}\log\rho_{BC})-\tr(\rho_{C}\log\rho_{C})
\]
is the \define{quantum conditional entropy} of $\rho_{BC}$ given $\rho_{C}$. 
One can show that 
\[
\RE\big((F\circ G,R\circ Q)\big)=H(AB|C)+\log(d_{A})+\log(d_{B}),
\]
\[
\RE\big((G,R)\big)=H(B|C)+\log(d_{B}),
\quad\text{and}\quad
\RE\big((F,Q)\big)=H(A|BC)+\log(d_{A})
\]
by applying Proposition~\ref{prop:somefunctoriality} to the composable pair
\[
\Big(\mC,\tr(\rho_{C}\;\cdot\;)\Big)\xrt{\left(G,R\right)}\Big(\mB\otimes\mC,\tr(\rho_{BC}\;\cdot\;)\Big)\xrt{\left(F,Q\right)}\Big(\mA\otimes\mB\otimes\mC,\tr(\rho_{ABC}\;\cdot\;)\Big),
\]
where $G$ and $F$ are the standard inclusions, $\upsilon_{B}:=\frac{1}{d_{B}}\mathds{1}_{d_{B}}$, $\upsilon_{A}:=\frac{1}{d_{A}}\mathds{1}_{d_{A}}$, and $R$ and $Q$ are the CPU maps given by 
$R:=\tr_{\mB}\left(\upsilon_{B}\otimes1_{\mC}\;\cdot\;\right)$,
and $Q:=\tr_{\mA}\left(\upsilon_{A}\otimes1_{\mB}\otimes1_{\mC}\;\cdot\;\right)$. This reproduces~(\ref{eq:qchainrule}). 
\end{example}

Proposition~\ref{prop:somefunctoriality} does not fully prove functoriality of $\RE$. One still needs to check functoriality in case one of the terms is infinite %
(eg.\ if $\relent{\omega}{\zeta\circ R\circ Q}=\infty$, then at least one of $\relent{\xi}{\zeta\circ R}$ or $\relent{\omega}{\xi\circ Q}$ must be infinite, and conversely). This will be addressed in future work. In the remainder, we prove affinity of $\RE$.

\begin{definition}
Given $\lambda\in[0,1]$, set $\ov\lambda:=1-\lambda$. The $\lambda$-weighted \define{convex sum} $\lambda(\mA,\omega)\oplus\ov\lambda(\ov\mA,\ov\omega)$ of objects $(\mA,\omega)$ and $(\ov\mA,\ov\omega)$ in $\NCFinStat$ is given by the pair $(\mA\oplus\ov\mA,\lambda\omega\oplus\ov\lambda\ov\omega)$, where 
$
\big(\lambda\omega\oplus\ov\lambda\ov\omega\big)(A\oplus \ov A):=\lambda\omega(A)+\ov\lambda\ov\omega(\ov A)$ whenever $A\in\mA,\,\ov A\in\ov\mA.$
The \define{convex sum} $\lambda(F,Q)\oplus\ov\lambda(\ov F,\ov Q)$ of $(\mB,\xi)\xrt{(F,Q)}(\mA,\omega)$ and $(\ov\mB,\ov\xi)\xrt{(\ov F,\ov Q)}(\ov\mA,\ov\omega)$ is the morphism $(F\oplus \ov F,Q\oplus \ov Q)$. A functor $\NCFinStat\xrightarrow{\mathfrak{L}}\B[0,\infty]$ is \define{affine} iff 
$
\mathfrak{L}\big(\lambda(F,Q)\oplus\ov\lambda(\ov F,\ov Q)\big)=\lambda \mathfrak{L}(F,Q)+\ov\lambda\mathfrak{L}(\ov F,\ov Q)
$
for all pairs of morphisms in $\NCFinStat$ and $\lambda\in[0,1]$. 
\end{definition}

\begin{proposition}
Let $(\mB,\xi)\xrt{(F,Q)}(\mA,\omega)$ and $(\ov\mB,\ov\xi)\xrt{(\ov F,\ov Q)}(\ov\mA,\ov\omega)$ be two morphisms for which $\RE(F,Q)$ and $\RE(\ov F,\ov Q)$ are finite. Then $\RE\big(\lambda(F,Q)\oplus\ov\lambda(\ov F,\ov Q)\big)=\lambda \RE(F,Q)+\ov\lambda\RE(\ov F,\ov Q)$. 
\end{proposition}

\bprf
When $\lambda\in\{0,1\}$, the claim follows from the convention $0\log0=0$. For $\lambda\in(0,1)$, temporarily set $\mu:=\RE\big(\lambda(F,Q)\oplus\ov\lambda(\ov F,\ov Q)\big)$.
Then 
\[
\begin{split}
\mu&\overset{\text{(\ref{eq:SomxQ})}}{=\joinrel=\joinrel=}\sum_{x\in X}\tr\Big(\lambda p_{x}\rho_{x}\log(\lambda p_{x}\rho_{x})\Big)+\sum_{\ov x\in\ov X}\tr\Big(\ov\lambda \ov p_{\ov x}\ov\rho_{\ov x}\log\big(\ov\lambda \ov p_{\ov x}\ov\rho_{\ov x}\big)\Big)\\
&-\sum_{x,y}\left[\tr\Big(\lambda p_{x}\rho_{x;yy}\big(\log(\alpha_{yx})\otimes\mathds{1}_{n_{y}}\big)\Big)+\tr\Big(\lambda p_{x}\tr_{\mM_{c^{F}_{yx}}}(\rho_{x;yy})\log(\lambda q_{y}\sigma_{y})\Big)\right]\\
&-\sum_{\ov x,\ov y}\left[\tr\Big(\ov\lambda \ov p_{\ov x}\rho_{\ov x;\ov y \ov y}\big(\log(\ov\alpha_{\ov y \ov x})\otimes\mathds{1}_{\ov n_{\ov y}}\big)\Big)
+\tr\Big(\ov\lambda \ov p_{\ov x}\tr_{\mM_{c^{\ov F}_{\ov y\ov x}}}(\rho_{\ov x;\ov y \ov y})\log\big(\ov\lambda \ov q_{\ov y}\ov\sigma_{\ov y}\big)\Big)\right],
\end{split}
\]
where we have used bars to denote analogous expressions for the algebras, morphisms, and states with bars over them. 
From this, the property $\log(ab)=\log(a)+\log(b)$ of logarithms is used to complete the proof. 
\eprf  

In summary, we have taken the first steps towards illustrating that the quantum relative entropy may have a functorial description along similar lines to those of the classical one in~\cite{BaFr14}.  
Using the recent non-commutative disintegration theorem~\cite{PaRu19}, we have proved parts of affinity and functoriality of the relative entropy.
The importance of functoriality comes from the connection between the quantum relative entropy and the reversibility of morphisms~\cite[Theorem~4]{Pe86}.
For example, optimal hypotheses are Bayesian inverses~\cite[Theorem~8.3]{PaBayes}, which admit stronger compositional properties~\cite[Propositions~7.18 and~7.21]{PaBayes} than alternative recovery maps in quantum information theory~\cite[Section~4]{Wilde15}.%
\footnote{One must assume faithfulness for some of the calculations in~\cite[Section~4]{Wilde15}. The compositional properties in~\cite[Proposition~7.21]{PaBayes}, however, need no such assumptions.}
In future work, we hope to prove functoriality (without any faithfulness assumptions), continuity, and a complete characterization.

\vspace{2mm}
\noindent
\textbf{Acknowledgements.} The author thanks the reviewers of GSI'21 for their numerous helpful suggestions. 
This research has also received funding from the European Research Council (ERC) under the European Union's Horizon 2020 research and innovation program (QUASIFT grant agreement 677368).

%

 \bibliographystyle{splncs04}
 \bibliography{relent}

\begin{thebibliography}{10}
\providecommand{\url}[1]{\texttt{#1}}
\providecommand{\urlprefix}{URL }
\providecommand{\doi}[1]{https://doi.org/#1}

\bibitem{BaFr14}
Baez, J.C., Fritz, T.: A {B}ayesian characterization of relative entropy.
  Theory Appl. Categ.  \textbf{29}(16),  422--457 (2014)

\bibitem{BFL}
Baez, J.C., Fritz, T., Leinster, T.: A characterization of entropy in terms of
  information loss. Entropy  \textbf{13}(11),  1945--1957 (2011).
  \doi{10.3390/e13111945}

\bibitem{Fa01}
Farenick, D.R.: Algebras of linear transformations. Universitext,
  Springer-Verlag, New York (2001). \doi{10.1007/978-1-4613-0097-7}

\bibitem{FuJa15}
Furber, R., Jacobs, B.: From {K}leisli categories to commutative
  {$C^*$}-algebras: probabilistic {G}elfand duality. Log. Methods Comput. Sci.
  \textbf{11}(2),  1:5, 28 (2015). \doi{10.2168/LMCS-11(2:5)2015}

\bibitem{GaPa18}
Gagn\'e, N., Panangaden, P.: A categorical characterization of relative entropy
  on standard {B}orel spaces. In: The {T}hirty-third {C}onference on the
  {M}athematical {F}oundations of {P}rogramming {S}emantics ({MFPS} {XXXIII}),
  Electron. Notes Theor. Comput. Sci., vol.~336, pp. 135--153. Elsevier Sci. B.
  V., Amsterdam (2018). \doi{10.1016/j.entcs.2018.03.020}

\bibitem{Pa17}
Parzygnat, A.J.: Discrete probabilistic and algebraic dynamics: a stochastic
  {G}elfand--{N}aimark theorem (2017), arXiv preprint:
  \href{https://arxiv.org/abs/1708.00091}{1708.00091 [math.FA]}

\bibitem{PaEntropy}
Parzygnat, A.J.: A functorial characterization of von~{N}eumann entropy (2020),
  arXiv preprint: \href{https://arxiv.org/abs/2009.07125}{2009.07125
  [quant-ph]}

\bibitem{PaBayes}
Parzygnat, A.J.: Inverses, disintegrations, and {B}ayesian inversion in quantum
  {M}arkov categories (2020), arXiv preprint:
  \href{https://arxiv.org/abs/2001.08375}{2001.08375 [quant-ph]}

\bibitem{PaRu19}
Parzygnat, A.J., Russo, B.P.: Non-commutative disintegrations: existence and
  uniqueness in finite dimensions (2019), arXiv preprint:
  \href{https://arxiv.org/abs/1907.09689}{1907.09689 [quant-ph]}

\bibitem{Pe86}
Petz, D.: Sufficient subalgebras and the relative entropy of states of a
  von~{N}eumann algebra. Commun. Math. Phys.  \textbf{105}(1),  123--131
  (1986). \doi{10.1007/BF01212345}

\bibitem{Pe92}
Petz, D.: Characterization of the relative entropy of states of matrix
  algebras. Acta Math. Hung.  \textbf{59}(3-4),  449--455 (1992).
  \doi{https://doi.org/10.1007/BF00050907}

\bibitem{Um62}
Umegaki, H.: Conditional expectation in an operator algebra. {IV}. entropy and
  information. Kodai Math. Sem. Rep.  \textbf{14}(2),  59--85 (1962).
  \doi{10.2996/kmj/1138844604}

\bibitem{Wilde15}
Wilde, M.M.: Recoverability in quantum information theory. Proc. R. Soc. A.
  \textbf{471},  20150338 (2015). \doi{10.1098/rspa.2015.0338}

\end{thebibliography}

%
%
%
%
\end{document}